%% file: main.tex
\documentclass[conference]{IEEEtran}
\IEEEoverridecommandlockouts

\usepackage{amsmath,graphicx}
\usepackage{multirow}
\usepackage{booktabs}
\usepackage{amsfonts}

\usepackage{bbding}
\usepackage[urlcolor=blue]{hyperref}
\usepackage{makecell} 
\usepackage{mathrsfs}
\usepackage{threeparttable}
\usepackage{adjustbox}
\usepackage{float}
\usepackage{caption}
\usepackage{graphicx}
\usepackage{float} 
\usepackage{graphicx}
\usepackage{float}
\usepackage{subfig}

\def\BibTeX{{\rm B\kern-.05em{\sc i\kern-.025em b}\kern-.08em
    T\kern-.1667em\lower.7ex\hbox{E}\kern-.125emX}}
\begin{document}

\title{SnakeGAN: A Universal Vocoder Leveraging DDSP Prior Knowledge and Periodic Inductive Bias\\
}
\makeatletter
\newcommand{\linebreakand}{%
  \end{@IEEEauthorhalign}
  \hfill\mbox{}\par
  \mbox{}\hfill\begin{@IEEEauthorhalign}
}
\makeatother

\author{
\IEEEauthorblockN{1\textsuperscript{st} Sipan Li*}
\IEEEauthorblockA{\textit{Shenzhen International Graduate School} \\
\textit{Tsinghua University}\\
Shenzhen, China \\
lsp20@mails.tsinghua.edu.cn}
\and
\IEEEauthorblockN{2\textsuperscript{nd} Songxiang Liu†}
\IEEEauthorblockA{\textit{AI Lab} \\
\textit{Tencent Inc}\\
Shenzhen, China \\
shaunxliu@tencent.com}
\and
\IEEEauthorblockN{3\textsuperscript{rd} Luwen Zhang}
\IEEEauthorblockA{\textit{Shenzhen International Graduate School} \\
\textit{Tsinghua University}\\
Shenzhen, China \\
zlw20@mails.tsinghua.edu.cn}
\linebreakand
\IEEEauthorblockN{4\textsuperscript{th} Xiang Li}
\IEEEauthorblockA{\textit{Shenzhen International Graduate School} \\
\textit{Tsinghua University}\\
Shenzhen, China \\
xiang-li20@mails.tsinghua.edu.cn}
\and
\IEEEauthorblockN{5\textsuperscript{th} Yanyao Bian}
\IEEEauthorblockA{\textit{AI Lab} \\
\textit{Tencent Inc}\\
Shenzhen, China \\
louisbian@tencent.com}
\and
\IEEEauthorblockN{6\textsuperscript{th} Chao Weng}
\IEEEauthorblockA{\textit{AI Lab} \\
\textit{Tencent Inc}\\
Shenzhen, China \\
cweng@tencent.com}
\linebreakand
\IEEEauthorblockN{7\textsuperscript{th} Zhiyong Wu†}
\IEEEauthorblockA{\textit{Shenzhen International Graduate School} \\
\textit{Tsinghua University}\\
Shenzhen, China \\
zywu@se.cuhk.edu.hk}
\and
\IEEEauthorblockN{8\textsuperscript{th} Helen Meng}
\IEEEauthorblockA{\textit{Department of Systems Engineering and Engineering Management} \\
\textit{The Chinese University of Hong Kong}\\
Hong Kong SAR, China \\
hmmeng@se.cuhk.edu.hk}
}

\maketitle
\renewcommand{\thefootnote}{\fnsymbol{footnote}} 
\footnotetext[1]{Work done during the internship at Tencent AI Lab}
\footnotetext[2]{Corresponding authors}
\begin{abstract}
\input{text/abstract}
\label{sec:abs}
\end{abstract}


\section{Introduction}
\input{text/introduction.tex}
\label{sec:intro}

\section{Related Woek}
\input{text/related_works}
\label{sec:related}

\section{Methodology}
\input{text/method.tex}
\label{sec:method}

\section{Experiments}
\input{text/exp.tex}
\label{sec:exp}

\section{Conclusion}
\input{text/conclusion.tex}

\label{sec:conclusion}

\section*{Acknowledgment}


This work is supported by the National Natural Science Foundation of China (62076144), Shenzhen Science and Technology Program (WDZC20220816140515001, JCYJ20220818101014030), Tencent AI Lab Rhino-Bird Focused Research Program (RBFR2022005) and Tsinghua University - Tencent Joint Laboratory.

\bibliographystyle{IEEEbib}
\bibliography{ref}
\end{document}

%% file: text/abstract.tex
Generative adversarial network (GAN)-based neural vocoders have been widely used in audio synthesis tasks due to their high generation quality, efficient inference, and small computation footprint.
However, it is still challenging to train a universal vocoder which can generalize well to out-of-domain (OOD) scenarios, such as unseen speaking styles, non-speech vocalization, singing, and musical pieces.
In this work, we propose SnakeGAN, a GAN-based universal vocoder, which can synthesize high-fidelity audio in various OOD scenarios.
SnakeGAN takes a coarse-grained signal generated by a differentiable digital signal processing (DDSP) model as prior knowledge, aiming at recovering high-fidelity waveform from a Mel-spectrogram. We introduce periodic nonlinearities through the Snake activation function and anti-aliased representation into the generator, which further brings desired inductive bias for audio synthesis and significantly improves the extrapolation capacity for universal vocoding in unseen scenarios.
To validate the effectiveness of our proposed method, we train SnakeGAN with only speech data and evaluate its performance for various OOD distributions with both subjective and objective metrics. 
Experimental results show that SnakeGAN significantly outperforms the compared approaches and can generate high-fidelity audio samples including unseen speakers with unseen styles, singing voices, instrumental pieces, and nonverbal vocalization.

%% file: text/introduction.tex
\begin{figure*}[t]
	\centering
    \includegraphics[width=0.85\linewidth]{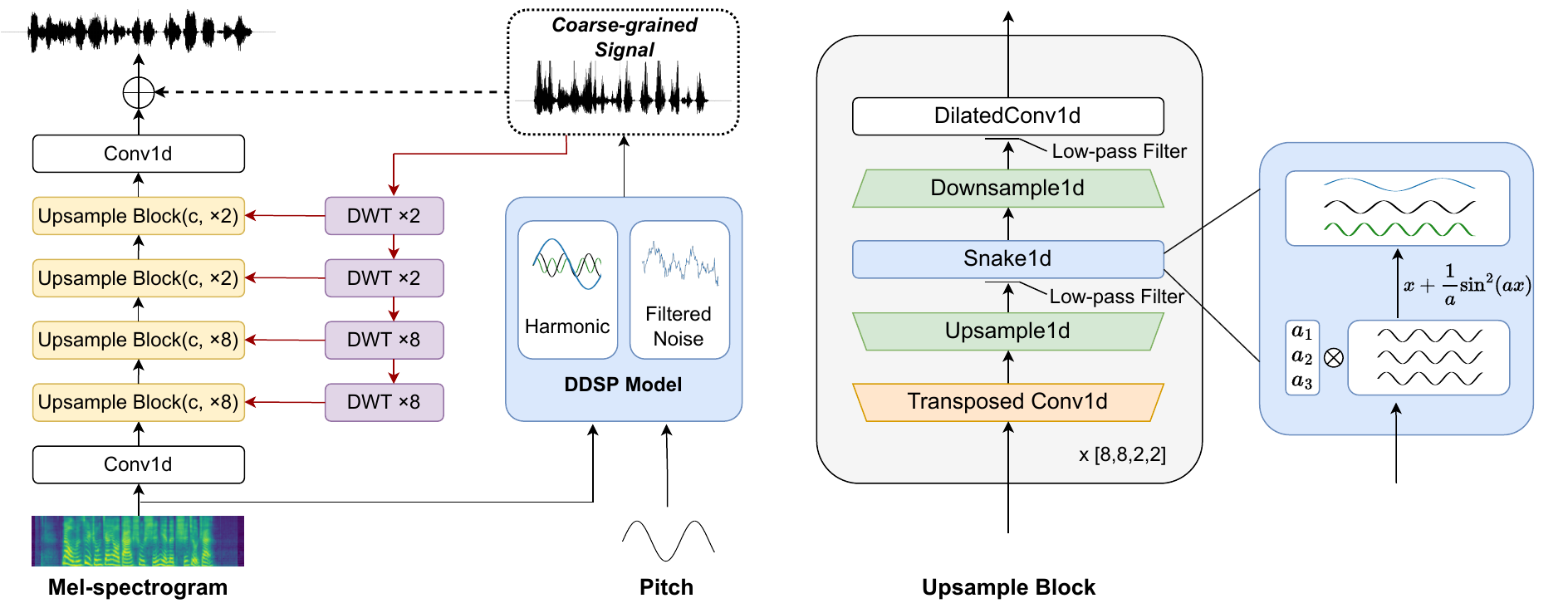}
	\caption{
 Schematic diagram of SnakeGAN generator. The generator is composed of multiple transposed-convolution-based upsampling blocks, where hidden features are enhanced by anti-aliased multi-periodicity composition modules.
 It applies the Snake activation function for periodic inductive bias and filtered nonlinearities for anti-aliasing purposes.
 The DDSP oscillator generates the coarse-grained signal as time-domain prior and then applied to the generator block after N times DWT downsample at each block. 
	}
	\label{fig:architecture}
	\vspace{-0.2cm}
\end{figure*}

%

Neural vocoders \cite{kong2020diffwave, nercessian2022differentiable, huang2022fastdiff} have drawn much attention as they generate the final waveform from acoustic information in many applications like Text-to-Speech (TTS) \cite{ren2021portaspeech}, singing voice synthesis\cite{chen2020hifisinger}, voice conversion \cite{kaneko2021maskcyclegan}, etc.
Most high-fidelity neural vocoders are based on the generative adversarial network (GAN) and have shown their advantages in generating raw waveform conditioned on Mel-spectrogram with fast inference speed and lightweight networks \cite{kong2020hifi,kaneko2022istftnet,lee2022bigvgan, kim2021fre, bak2022avocodo}.
Existing works on GAN-based neural vocoders mainly focus on improving the discriminator architecture\cite{you2021gan} or incorporating auxiliary training losses into the adversarial training.
MelGAN\cite{kumar2019melgan} first realizes a competitive GAN network vocoder by introducing a multi-scale discriminator (MSD) that downsamples the raw waveform at multiple scales through average pooling, leading to a loss of high-frequency information.
Parallel WaveGAN\cite{yamamoto2020parallel} improves the training loss by extending the short-time Fourier transform (STFT) loss to be multi-resolution. 
Multi-period discriminator (MPD) and multi-receptive field fusion (MRF) are proposed by HiFi-GAN\cite{kong2020hifi}, which achieves high-fidelity performance. 

In real applications, however, neural vocoders typically suffer from heavy quality degradation when directly applied to unseen data.
It is of significant meaning to achieve the flexible generation of high-quality audio under various scenarios without any fine-tuning.
Therefore, the universal vocoders aim to improve the ability to model the robust mapping between the condition and the target (e.g. Mel-spectrogram and waveform), especially on the out-of-domain (OOD) inference data.

Recent works on universal vocoders like Universal MelGAN \cite{jang2020universal} and UnivNet \cite{jang2021univnet} utilize the multi-resolution discriminator (MRD) to enhance model generalization on OOD data,
which takes the multi-resolution spectrograms as the input and sharpens the spectral structure of the generated waveforms.
Nevertheless, the existing works still conduct OOD robustness tests on speech data, concentrating on unseen speakers and unseen languages.
It's still challenging to build a universal vocoder for various scenarios with a larger gap between the speech training data, such as the singing voice, instrumental pieces, and nonverbal vocalization.


For the purpose of further enhancing the effectiveness and robustness of the generator, we propose a universal neural vocoder named SnakeGAN.
SnakeGAN improves the waveform generator by introducing both the DDSP-based prior knowledge of waveform composition,
and the periodic non-linearities through incorporating the Snake activation function \cite{ziyin2020neural}.
Specifically, the Snake generator first obtains a coarse-grained DDSP-generated signal waveform and downsamples it for N times by Discrete Wavelet Transform (DWT)\cite{kim2021fre}, which can keep the high-frequency component.
Then, each of the downsampled signals is added to the corresponding upsample block, which is composed of the transposed convolutional block, followed by the anti-aliased multi-periodicity composition module with Snake activation.
On the one hand,
the snake activation function achieves the desired periodic inductive bias to learn a periodic function while maintaining a favorable optimization property of the ReLU-based activations.
On the other hand,
coupling the characteristics of the time-domain periodic and aperiodic components prior provided by DDSP strengthens the generator's robustness under unseen scenarios. 

We choose speech, singing voice, instrumental pieces, and nonverbal vocalization as the target scenarios in which vocoders are mainly used. In our experiments, the proposed model generates high-quality audio in various scenarios, outperforming the state-of-the-art DDSP-based vocoder and the mainstream HiFi-GAN vocoder.
The audio synthesized by the proposed universal SnakeGAN vocoder and other models is available at our demo page\footnote{Demo page: \href{https://github.com/thuhcsi/icme2023-snakeGAN-vocoder/}{https://github.com/thuhcsi/SnakeGAN/}}.

In general, the contributions of this paper are three-fold: 
\begin{itemize}

\item We demonstrate by experiments that DDSP-based vocoders have better robustness when given a small amount of data.

\item We introduce the state-of-the-art generator and discriminator with the periodic inductive Snake activation function, which can highly eliminate aliasing artifacts and improves audio quality. 

\item We propose a novel and effective GAN vocoder, which can generalize well to universal scenarios by conditioned on DDSP prior even with a large amount of data.

\end{itemize}

%% file: text/related_works.tex
\subsection{Preliminaries of typical GAN vocoder}
\label{sec:typical GAN}

GAN-based vocoder generates waveform normally by a few transposed convolution upsampling network layers which also contain a stack of residual blocks with dilated convolutions. Typically, multiple discriminators are adopted for adversarial training to learn different frequency domain features of audio. 

\subsubsection{Generator}
Specifically, to address the problem of generalization ability, BigVGAN\cite{lee2022bigvgan} proposes the anti-aliased multi-periodicity composition (AMP) block with Snake activation function\cite{ziyin2020neural}. The BigVGAN's generator with AMP block is similar to the structure of StyleGAN3\cite{karras2021alias}, which has shown satisfying generalization ability in the image generation domain.  

Meanwhile, the Snake activation function, defined as $f(x) = x + sin^2(x)$, is demonstrated in\cite{ziyin2020neural} that can bring periodic inductive bias and can perform well for temperature and financial data prediction. Considering the audio waveform is known to exhibit high periodicity and can be represented as a composition of primitive periodic components, BigVGAN suggests that we can provide the desired inductive bias to the generator architecture based on Snake.

In addition, StyleGAN3\cite{karras2021alias} identifies that the aliasing artifacts in image synthesis are rooted in careless signal processing. StyleGAN3 applies the nonlinearity to the temporarily increased resolution (e.g. 2×) that approximates the continuous representation inspired by the Nyquist-Shannon sampling theorem. The continuous representation of nonlinearity ensures translation equivariance in the feature space, and the nonlinearity generates novel frequencies in the continuous domain, thereby eliminating the aliasing.

\subsubsection{Discriminator}
The state-of-the-art GAN vocoders usually comprise several types of discriminators to guide the generator to synthesize coherent waveform while minimizing perceptual artifacts which are easily detectable. We apply the Fre-GAN's setting of discriminators, including MPD and MSD, both with DWT instead of average pooling. Noteworthy, the average pooling ignores the sampling theorem, and high-frequency contents are aliased and become invalid, while DWT is an efficient but effective way of downsampling non-stationary signals into several frequency sub-bands and can preserve high-frequency components better. A few recent works propose to apply the discriminator on the time–frequency domain using the multi-resolution discriminator (MRD). MRD is also composed of several subdiscriminators that operate on multiple 2-D linear spectrograms with different STFT resolutions. We also apply MRD to improve the quality by sharpening the signal in the spectral domain with reduced pitch and periodicity artifacts.
%

\subsection{Overview of DDSP}
\label{sec:DDSP}
The DDSP\cite{engel2020ddsp} model\footnote{\href{https://github.com/acids-ircam/ddsp_pytorch}{https://github.com/acids-ircam/ddsp\_pytorch}} has shown the ability to decouple and further control the characters of a time domain waveform.
It can flexibly adjust the amplitude, envelope, and fundamental frequency of audio respectively, and then decode these characters into the harmonic structure and filtered noise, which can precisely meet our goal to simulate the prior knowledge of target audio from different domains.
%
%
%

According to HNM, audio signal $s(t)$ can be represented as the sum of the harmonic $s_h(t)$ and noise components $s_n(t)$:
\begin{equation}
    s(t) = s_h(t) + s_n(t).
\end{equation}
For the voiced part, the signal can be approximated by superimposing a series of harmonic components whose pitches are the integer multiples of the fundamental frequency:
\begin{equation}
    s_h(t) = \sum_{k=-L(t)}^{L(t)}A_k(t)e^{jk\omega_0(t)t},
\end{equation}
in which $L(t)$ denotes the number of harmonic.
$A_k(t)$ denotes the amplitude and $\omega_0(t)$ denotes the fundamental frequency.
And for the unvoiced part, the signal can be directly represented by random noise based on the time-varying auto-regressive (AR) model $h(\tau,t)$:
\begin{equation}
    s_n(t) = e(t)[h(\tau,t)*b(t)],
\end{equation}
where $e(t)$ denotes the spectral envelope of noise signal and $b(t)$ denotes white noise signal.

%% file: text/method.tex

This section is organized as follows:
Section \ref{sec:pipeline} will introduce the overall pipeline of the proposed model architecture.
%
Section \ref{sec:DDSP} introduces the DDSP model and proposes a novel method that uses the Snake-based upsampling blocks to introduce periodic inductive bias and time-varying harmonic-plus-noise prior knowledge, making the generator perform better in extrapolation.

\subsection{Overall pipeline}
\label{sec:pipeline}

The pipeline of our proposed model architecture for the universal vocoder is represented in Figure \ref{fig:architecture}.

It consists of two main stages.
To begin with, based on the prior knowledge from different target audio domains, including speech, singing voice, instrumental pieces, and nonverbal vocalization, we model the distributions of acoustic features corresponding to fundamental frequency $f_0$, harmonic distribution $D$, harmonic amplitude $A$, and time-varying filtered noise through DDSP, a typical Harmonic-plus-Noise Model (HNM). 

Next, we propose two versions of the SnakeGAN generator, the SnakeGANv1 is the dotted line while the SnakeGANv2 is the solid red line, as is shown in Figure\ref{fig:architecture}. 

Lastly, we refer to Fre-GAN\cite{kim2021fre} and UnivNet\cite{jang2021univnet}, MPD with DWT and MRD discriminators are adopted to improve the synthesized audio quality.

%
\subsection{Introducing DDSP prior into black-box GAN}
\label{sec:DDSP}

The DDSP module in our work generates a coarse-grained signal in the time domain. We note that the DDSP signal is generated combining prior knowledge from both harmonic oscillator and filtered noise, and the black-box mode GAN generator is lack of such guidance in the time domain. 
The natural way to think about it would be how to introduce the prior into the generator to make it more robust. 
The SnakeGANv1 simply adds the DDSP signal to the synthesized audio, it is a simple but effective way. 
Additionally, we present SnakeGANv2. The SnakeGANv2 generator aims to couple the time-domain signal of DDSP with the GAN generator more effectively and combines with the Snake activation function. 
we downsample the DDSP signal N times by DWT, as the up sample multiple of the generator is [8, 8, 2, 2], thus the DWT down sample multiple is [2, 2, 8, 8], correspondingly. 
At last, we add the down-sampled signals to each upsampling block respectively as time-domain supervision to guide the Snake generator learning.

%% file: text/exp.tex

\begin{table*}[!htbp]
\centering
\caption{Subjective evaluation results (MOS values). ``SnakeGANv1" denotes the method that simply adds the DDSP signal to the generator. ``SnakeGANv2" denotes the approach that couples the signals after DWT down-sampled with each upsample block.  ``CI'' denotes the confidence interval. $\dagger$ denotes the proposed vocoder.}
\label{tab:mos}
\resizebox{0.9\linewidth}{!}
{
\begin{tabular}{ccccccccc}
\toprule
\multirow{2}{*}{Model} & \multicolumn{2}{c}{\shortstack{unseen styles \\ (OOD-expressive)}} & \multicolumn{2}{c}{\shortstack{singing\\voices}} & \multicolumn{2}{c}{\shortstack{instrumental \\ pieces}} & \multicolumn{2}{c}{\shortstack{nonverbal \\ vocalization}} \\
\cmidrule(lr){2-3} \cmidrule(lr){4-5} \cmidrule(lr){6-7} \cmidrule(lr){8-9}
                            & MOS$\uparrow$      & 95\% CI                     & MOS$\uparrow$           & 95\% CI           & MOS$\uparrow$          & 95\% CI          & MOS$\uparrow$           & 95\% CI           \\ \hline
Ground Truth                & 4.64     & $\pm$ 0.07     & 4.86     & $\pm$ 0.05    & 4.76      & $\pm$ 0.08      & 4.42        & $\pm$ 0.10           \\ \hline
HiFi-GAN (V1)               & 4.12     & $\pm$ 0.09     & 3.52     & $\pm$ 0.09       & 3.18             & $\pm$ 0.10                 &  3.66              & $\pm$ 0.12           \\
HooliGAN          & 4.07     & $\pm$ 0.08     & 3.36              & $\pm$ 0.10        & 2.97             & $\pm$ 0.10                 & 3.40              & $\pm$ 0.11        \\
SnakeGANv1          & \textbf{4.37}     & $\pm$ 0.08     & 3.44              & $\pm$ 0.09        & 3.16            & $\pm$ 0.10                 & \textbf{3.78}              & $\pm$ 0.12        \\
SnakeGANv2 $^\dagger$        & \textbf{4.39}     & $\pm$ 0.08     & \textbf{3.70}      & $\pm$ 0.09        & \textbf{3.34}             & $\pm$ 0.10                 & \textbf{3.89}        & $\pm$ 0.12 \\              
\bottomrule 
\end{tabular}
}
\end{table*}

\begin{table}[!htbp]
\centering
\caption{Objective evaluation results (PESQ, STOI, and MR-STFT Loss values) of unseen styles and singing voices.}

\label{tab:pesqandstoi}
\resizebox{\linewidth}{!}{
\begin{tabular}{cccc}
\toprule
\multirow{2}{*}{Metric} & \multirow{2}{*}{Model}                  & \multirow{2}{*}{\shortstack{unseen styles \\ (OOD-expressive)}} & \multirow{2}{*}{\shortstack{singing\\voices}}\\ 
 & & & \\
\midrule
\multirow{4}{*}{PESQ$\uparrow$}      & HiFi-GAN (V1)                    & 3.059                  & 2.785                                                                                            \\
                           & HooliGAN            & 2.916                   & 2.594                                                                                  \\ 
                           & SnakeGANv1        & \textbf{3.289}           & \textbf{2.848}                                                                       \\ 
                           & SnakeGANv2$^\dagger$      & \textbf{3.264}              & 2.642                                                                        \\ \midrule
\multirow{4}{*}{STOI$\uparrow$}      & HiFi-GAN (V1)                & 0.968                     & \textbf{0.845}                                                                                             \\
                           & HooliGAN           & 0.954                     &  0.816                                                                                           \\ 
                           & SnakeGANv1        & \textbf{0.972}            & 0.823                                                                                      \\ 
                           & SnakeGANv2$^\dagger$       & \textbf{0.972}             & 0.823                                                                                        
                                          \\ \midrule
\multirow{4}{*}{MR-STFT Loss$\downarrow$}      & HiFi-GAN (V1)                & 1.020                     & 1.329                                                                                             \\
                           & HooliGAN           & 1.074                     & 1.344                                                                                            \\ 
                           & SnakeGANv1       & \textbf{0.987}            & 1.311                                                                                      \\ 
                           & SnakeGANv2$^\dagger$       & \textbf{0.985}             & \textbf{1.270}                                                                                         \\
\bottomrule
\end{tabular}}
\end{table}

To validate the effectiveness of our proposed method, we train SnakeGAN with only speech data and evaluate its performance with both subjective and objective metrics for various OOD distributions, including singing voice, instrumental pieces, and nonverbal vocalization.

\subsection{Corpus and data configuration}
The audio sample rate is 24KHz and the 80-band log Mel-spectrogram is extracted with a 1024-point FFT, 256 sample frameshift, and 1024 sample frame length. 


\subsubsection{Training set}
We use an internal gender-balanced multi-speaker speech corpus for training. The dataset contains 291 speakers and has duration of 278 hours in total. Most sentences are in Mandarin Chinese and the remaining sentences are in English or Chinese-English code-switched.

\subsubsection{Testing set}
We consider the following OOD scenarios in the test set:
\begin{itemize}

\item Unseen speakers with OOD-expressive styles

The unseen speakers with OOD-expressive styles data contain 1024 utterances and 8 speakers, and every utterance is highly expressive.

\item Singing voice

We evaluated our method on singing voice clips extracted from the Mandarin singing corpus dataset Opencpop\cite{wang2022opencpop}, which usually includes some skills such as trill, long tone, and leaning tone, which usually do not exist in speech.

\item Instrumental pieces

Audio clips were extracted from the single instrument musical pieces of URMP dataset\cite{10.1109/TMM.2018.2856090}. The URMP dataset is made of 44 simple multi-instrument music works, which are composed of performances recorded separately by a single track.

\item Nonverbal vocalization

We extracted audio clips from the Nonverbal Vocalization dataset\cite{denham2013beyond}, which is a human nonverbal vocal sound dataset containing crying, laughing, sneezing, moaning, screaming, etc.

\end{itemize}

\begin{table}[!htbp]
\centering
\vspace{-0.1em}
\caption{Pitch distribution of each dataset.} 
\label{tab:f0dist}
\resizebox{\linewidth}{!}
{
\begin{tabular}{cccccc}
\toprule
Dataset & \#Utterance & Min & Max & Mean & Std \\
\midrule
\multirow{1}{*}\makecell{Training set}  & 218k & -0.947                    & 6.258                  & 123.992                 & 130.806                                                                              \\
                           
\multirow{1}{*}\makecell{Test OOD styles}   &  1024 & -0.857                    & 7.477                  & 95.864                 & 111.842                                                                              \\
                           
\multirow{1}{*}\makecell{Singing }   & 300  & -1.554                    & 2.631                  & 277.634                 & 178.607                                                                \\
                           
\multirow{1}{*}\makecell{Instrumental} & 300 & -1.286                    & 2.625                  & 287.254                 & 223.335                                                                \\
                           
\multirow{1}{*}\makecell{Nonverbal} & 300 & -0.891                    & 3.391                  & 175.025                 & 196.476                                                                \\
                           
\bottomrule
\end{tabular}}
\vspace{-0.2cm}
\end{table}

Pitch features of each dataset are extracted from the ground-truth audio by praat-parselmouth\cite{jadoul2018introducing}. 
The pitch range after z-score normalization as well as the mean and standard deviation are shown in Table \ref{tab:f0dist}.
Much differences can be observed among different datasets.

\subsection{Investigation of the effectiveness of DDSP structure}
\label{sec:DDSP eff}

In this section, we refer to DDSP primarily concerning the additive oscillator and the model’s ability to learn time-varying amplitude envelopes. The generalization ability of the DDSP structure is investigated by implementing a DDSP-based vocoder HooliGAN\cite{mccarthy2020hooligan}, to validate the robustness of the DDSP architecture. 

\begin{figure}[t]
	\centering
    \includegraphics[width=\linewidth]{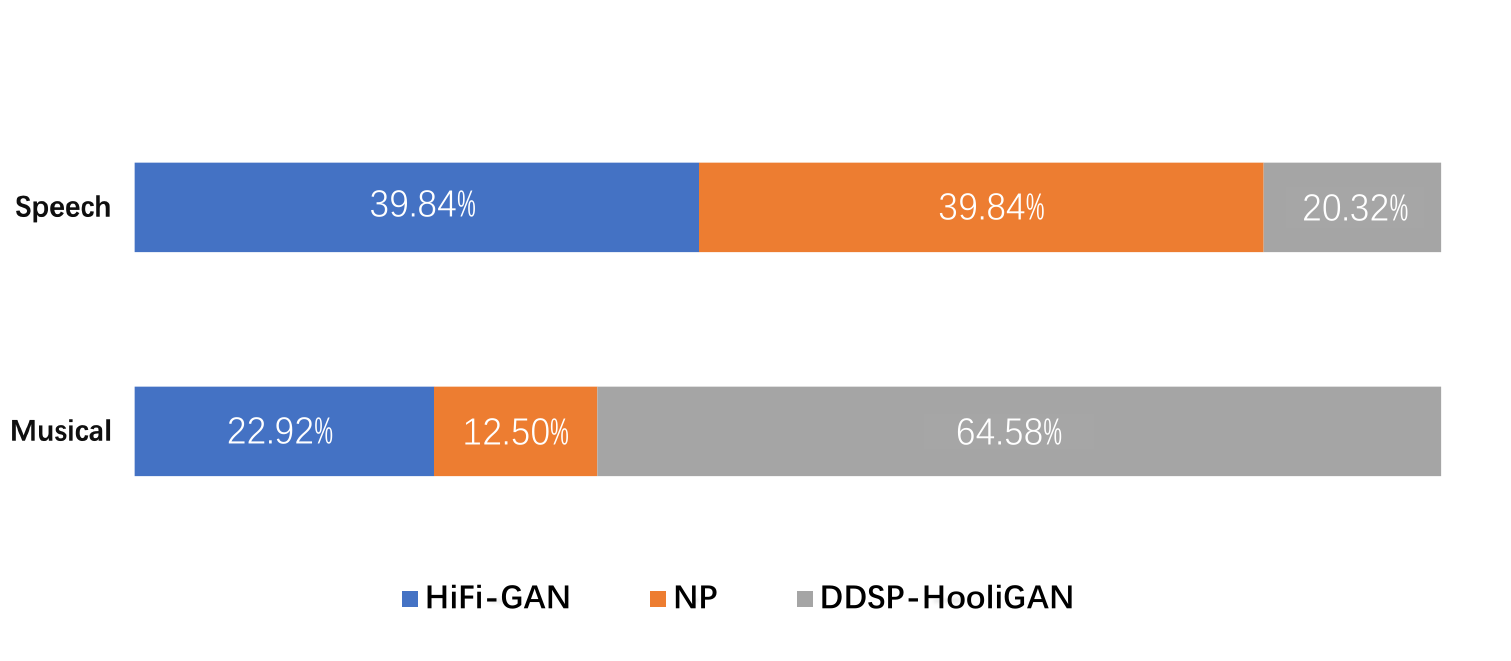}
    \vspace{-0.75cm}
	\caption{
	Results of ABX tests comparing DDSP-HooliGAN and HiFi-GAN on speech and musical pieces respectively.
	}
	\label{fig:ABX}
	\vspace{-0.2cm}
\end{figure}

We train a modified HooliGAN on LJSpeech dataset\cite{ljspeech17}, as the official open-source HiFi-GAN\footnote{\href{https://github.com/jik876/hifi-gan}{https://github.com/jik876/hifi-gan}} is trained on LJSpeech. Hereafter, to verify the robustness of DDSP structure, we compare the HooliGAN and HiFi-GAN with OOD scenarios, including unseen speakers and musical pieces. We conduct the ABX test to demonstrate the effectiveness of the DDSP structure, as it reasonably states the sound-generating mechanism. The results are shown in Fig.\ref{fig:ABX}. 64.58\%  participants selected HooliGAN on musical pieces and 22.92\% for HiFi-GAN. For speech utterances, 39.84\%  participants selected HiFi-GAN while 39.84\% selected NP, and 20.32\% selected HooliGAN. 
The results show that when feed with fewer data, DDSP-based HooliGAN can be more robust to some unseen scenarios, but HiFi-GAN is better when faced with speech.

It should be noted that
DDSP-based vocoders are usually small and have fewer parameters. 
Although they can perform well with only a small amount of data and are easy to train, it is still challenging and may lead to inferior audio quality with massive data. 
Experiments show that 
when using the 278-hour training set (Section 3.1.1), the generalization ability of HiFi-GAN exceeds that of HooliGAN.
Therefore, we decide not to take DDSP as generator directly. Instead, we introduce the DDSP coarse-grained signal as time-domain supervision to strengthen the state-of-the-art neural generator.

\vspace{-0.2cm}


\subsection{Experimental results of the candidate vocoders}
\label{sec:res}

\subsubsection{Speech}
We evaluated vocoders on several dimensions, including a subjective metric, mean opinion score (MOS) of audio quality (Table \ref{tab:mos}), and two objective metrics, perceptual evaluation of speech quality (PESQ) \cite{rix2001perceptual} and short-term objective intelligibility (STOI) \cite{taal2011algorithm} (Table \ref{tab:pesqandstoi}).
For each evaluation, we selected ten clips for the MOS test and three hundred clips for the PESQ and STOI tests, and also computed the Multi-Resolution STFT Loss. A total of twenty people participated in the MOS test.

For speech with OOD-expressive styles, the proposed SnakeGANv1 and SnakeGANv2 vocoders achieved the MOS score of 4.37 and 4.39, PESQ score of 3.289 and 3.264, STOI score of 0.972, and MR-STFT Loss 0.987 and 0.985, respectively.

Overall, experimental results on speeches proved the effectiveness of the proposed approach to introduce the time-domain DDSP signals as prior knowledge guidance and the effectiveness of the Snake activation function to strengthen the ability of generalization.


\subsubsection{Singing voice}
Since there is a big difference between the singing voice and speech, the vocoder trained on speech may degrade when facing the singing voice during inference. The results show that the proposed SnakeGANv2 achieved superior performance, with a 3.70 MOS score and 1.270 MR-STFT Loss.

\subsubsection{Instrumental pieces \& nonverbal vocalization}
Similar to the singing voice, instrumental pieces and nonverbal vocalization's distribution are various from speech, and without semantic information. Thus, we only refer to MR-STFT Loss as the metric, which is shown in Table\ref{tab:i&n}. The proposed SnakeGANv2 performs best of all models, which achieved 1.214 MR-STFT Loss, 3.34 MOS in instrumental pieces, and 1.242 MR-STFT Loss, 3.89 MOS in nonverbal vocalization.

\begin{table}[!htbp]
\centering
\caption{MR-STFT Loss values of instrumental pieces \& nonverbal vocalization.}

\label{tab:i&n}
\resizebox{\linewidth}{!}{
\begin{tabular}{cccc}
\toprule
\multirow{2}{*}{Metric} & \multirow{2}{*}{Model}                  & \multirow{2}{*}{\shortstack{instrumental \\ pieces }} & \multirow{2}{*}{\shortstack{nonverbal\\vocalization}}\\ 
 & & & \\
\midrule
\multirow{4}{*}{MR-STFT Loss$\downarrow$}      & HiFi-GAN (v1)               & 1.224                     & 1.326                                                                                             \\
                           & HooliGAN           &  1.287                     & 1.425                                                                                            \\ 
                           & SnakeGANv1        & 1.225            & 1.250                                                                                      \\ 
                           & SnakeGANv2$^\dagger$       & \textbf{1.214}             & \textbf{1.242}                                                                                         \\
\bottomrule
\end{tabular}}
\end{table}

%% file: text/conclusion.tex

In this paper, to improve the robustness of universal neural vocoding across diverse scenarios, and especially out-of-domain data, we present two versions of SnakeGAN. 
Specifically, we model the distributions of acoustic features under prior audio knowledge from multiple target scenarios through a DDSP module, the prior knowledge is then used as time-domain supervision to guide the GAN generator. 
The generalization of periodic components is explicitly modeled through the Snake activation function. 
In conclusion, a robust Snake generator and discriminator are applied in this work. 
Experimental results show that the proposed vocoder trained with a 278-hour speech corpus can be employed well and has achieved superior performance in many diverse scenarios.